\documentclass{amsart}

\usepackage{graphics}

\newcommand{\mathsym}[1]{{}}

\usepackage{amssymb,amscd}
\usepackage{amsmath}
\usepackage{amsthm}

\newtheorem{theorem}{Theorem}

\numberwithin{theorem}{section}

\theoremstyle{definition}

\newtheorem{example}[theorem]{Example}

\theoremstyle{remark}

\numberwithin{equation}{section}
\numberwithin{figure}{section}

\newfont{\germ}{eufm10}

\newcommand{\Z}{{\mathbb Z}}
\newcommand{\R}{{\mathbb R}}
\newcommand{\I}{{\mathbb I}}
\newcommand{\tI}{{\tilde {\mathbb I}}}

\begin{document}

\begin{center}
\begin{center}
\textbf{\Large{Combinatorial Bethe ansatz}\\
\Large{and ultradiscrete Riemann theta function}\\
\Large{with rational characteristics}
}
\end{center}

\vspace{6mm}
{\large Atsuo Kuniba and Reiho Sakamoto}
\end{center}

\vspace{0.2cm}
\begin{quotation}
{\small
A}{\tiny BSTRACT.}
{\small
The $U_q(\widehat{sl}_2)$ vertex model at $q=0$
with periodic boundary condition 
is an integrable cellular automaton in one-dimension.
By the combinatorial Bethe ansatz, 
the initial value problem is solved 
for arbitrary states in terms of an ultradiscrete analogue of the
Riemann theta function with rational characteristics.}
\end{quotation}

\noindent

\section{Introduction}\label{sec:intro}
Solvable vertex models \cite{Ba} associated with a
quantum affine algebra $U_q$ 
yield completely integrable one-dimensional cellular automata 
at $q=0$ \cite{HKT, HHIKTT, FOY}. 
The simplest case of $U_q(\widehat{sl}_2)$ with 
periodic boundary condition is known as 
the periodic box-ball system \cite{KTT,YYT}.
Here is an example of the time evolution pattern:

\begin{center}
$t=0$:\quad  111222221111222222221111111111111112222111211 \\
$t=1$:\quad  111111112222111111112222222221111111111222122 \\
$t=2$:\quad  222211111111222211111111111112222222221111211 \\
$t=3$:\quad  111122222222111122221111111111111111112222122 \\
$t=4$:\quad  222211111111222211112222222221111111111111211 \\
$t=5$:\quad  111122221111111122221111111112222222221111121 \\
$t=6$:\quad  222211112222111111112222111111111111112222212 \\
$t=7$:\quad  111122221111222222221111222221111111111111121 \\
$t=8$:\quad  111111112222111111112222111112222222221111112 \\
$t=9$:\quad  222211111111222211111111222211111111112222221 \\
\end{center}

\noindent
The dynamics is described as a motion of balls (letter $2$) hopping to the right 
along the periodical array of boxes.
The fusion transfer matrices $T_1, T_2, \ldots$ 
in the $q=0$ vertex model give rise to a  
commuting family of deterministic time evolutions.
The above example is the evolution under $T_9$. 

In \cite{KTT}, the initial value problem 
of the periodic box-ball system is solved by an 
inverse scattering method.
It synthesizes the combinatorial versions of the Bethe ansatz \cite{Be} 
at $q=1$ \cite{KR} and $q=0$ \cite{KN}.
The action-angle variables are constructed from 
the rigged configurations ($q=1$) by imposing a certain 
equivalence relations specified by the string center equation 
($q=0$).
The time evolution $T^t_l(p)$ of any state $p$ is 
obtained by an explicit algorithm 
whose computational steps are independent of $t$.

The Bethe ansatz approach \cite{KTT} captures several 
characteristic features in the quasi-periodic solutions 
of soliton equations \cite{DT,DMN,KM}.
It generalizes the 
connection between the original box-ball system 
on infinite lattice \cite{TS}
and soliton solutions via the ultradiscretization \cite{TTMS}
to a periodic setting.
For instance, the nonlinear dynamics of the periodic box-ball system 
becomes a straight motion of the Bethe roots (angle variable)
which live in an ultradiscrete analogue 
(\ref{eq:jv}) of the Jacobi variety.

In this paper we express the algorithmic solution of the 
initial value problem \cite{KTT} by an explicit formula
involving the ultradiscretization  
of the Riemann theta function with rational characteristics 
(${\bf z} \in \R^g, \,{\bf a} \in ({\mathbb Q}/\Z)^g)$:
\begin{equation}\label{eq:urt}
\begin{split}
\Theta_{\bf a} ({\bf z})&= \lim_{\epsilon \rightarrow +0}
\epsilon\log \vartheta_{\bf a}({\bf z})\\
&= -\min_{{\bf n} \in \Z^g}
\{{}^t({\bf n}+{\bf a})\Omega({\bf n}+{\bf a})/2
+{}^t({\bf n}+{\bf a}){\bf z}\}.
\end{split}
\end{equation}
Here $\vartheta_{\bf a}({\bf z})$ 
is the Riemann theta function \cite{M}
with a pure imaginary period matrix:
\begin{equation}\label{eq:vt}
\vartheta_{\bf a}({\bf z}) = \sum_{{\bf n} \in \Z^g}
\exp\Bigl(-\frac{{}^t({\bf n}+{\bf a})\Omega({\bf n}+{\bf a})/2
+{}^t({\bf n}+{\bf a}){\bf z}}{\epsilon}
\Bigr).
\end{equation}
Here $\Omega$ is a symmetric positive definite 
integer matrix (\ref{eq:omega}).
The function $\Theta_{\bf a} ({\bf z})$ enjoys the quasi-periodicity 
analogous to the Riemann theta function:
\begin{equation}\label{eq:qp}
\Theta_{\bf a}({\bf z}+{\bf v}) = 
\Theta_{\bf a}({\bf z})+{}^t{\bf v}\Omega^{-1}({\bf z}+{\bf v}/2)\quad 
\hbox{for any } \; {\bf v} \in 
\Omega\Z^g.
\end{equation}

Our main Theorem \ref{th:ima} covers {\em all} the states 
in the periodic box-ball system, which generalizes the 
earlier result \cite{KS} on those states 
in which the amplitudes of solitons were assumed to be distinct.
The theorem is derived from the recently obtained 
piecewise linear formula \cite{KSY} for 
the Kerov-Kirillov-Reshetikhin (KKR) bijection \cite{KR}
from rigged configurations to highest paths. 
It involves the ultradiscretization of the 
tau function for the KP hierarchy \cite{JM}.
A state $p$ of the 
periodic box-ball system is effectively treated as 
the infinite system  
$p \otimes p \otimes p \otimes \cdots$, and 
the ultradiscrete tau function for such states
turns out to be expressible in terms of 
${\Theta}_{\bf a}({\bf z})$ as in (\ref{eq:tau}).

In section \ref{sec:pbbs} we define 
the periodic box-ball system.
In section \ref{sec:cb} various data in 
the combinatorial Bethe ansatz at $q=1$ \cite{KR} 
and $q=0$ \cite{KN} are introduced.
In section \ref{sec:ist} we recall the 
solution of the initial value problem 
by the inverse scattering transform.
In section \ref{sec:ivp} we present our main 
Theorem \ref{th:ima} in this paper.

\section{Periodic box-ball system}\label{sec:pbbs}
Let us recall the periodic box-ball system without 
getting much into the crystal base theory \cite{Ka}.
For a positive integer $l$, let 
$B_l =\{(x_1,x_2) \in (\Z_{\ge 0})^2 \mid x_1+x_2=l\}$ and 
set $u_l = (l,0) \in B_l$. 
The element $(x_1,x_2)$ will also be denoted by 
the sequence 
$\overbrace{1\ldots 1}^{x_1}\overbrace{2\ldots 2}^{x_2}$, which 
is the semistandard Young tableau (without a frame) containing 
the letter $i$ $x_i$ times.
In this notation 
$u_l = \overbrace{1\ldots 1}^l$, and 
the two elements $(1,0)$ and $(0,1)$ in $B_1$ 
are simply written as $1$ and $2$, respectively.
In the following, the symbol $\otimes$ meaning the tensor product 
of crystals can just be understood as a product of sets.
Define the map $R: B_l \otimes B_1 \rightarrow B_1 \otimes B_l$ by
\begin{align*}
(x_1,x_2)\otimes 1 &\mapsto 
\begin{cases}1 \otimes (l,0) &\hbox{if } (x_1,x_2)=(l,0)\\
2 \otimes (x_1+1,x_2-1) & \hbox{otherwise},
\end{cases}\\
(x_1,x_2)\otimes 2 &\mapsto 
\begin{cases}2 \otimes (0,l) &\hbox{if } (x_1,x_2)=(0,l)\\
1 \otimes (x_1-1,x_2+1) & \hbox{otherwise}.
\end{cases}
\end{align*}
$R$ is a bijection and called the combinatorial $R$.
We write the relation $R(u \otimes b)= b' \otimes u'$ 
simply as $u \otimes b \simeq b' \otimes u'$, and similarly for 
any consequent relation of the form
$a\otimes u \otimes b \otimes c \simeq
a \otimes b' \otimes u' \otimes c$.

A state of the periodic box-ball system 
is an array of $1$ and $2$, which is 
regarded as an element 
$b_1 \otimes \cdots \otimes b_L \in B_1^{\otimes L}$
with $L$ being the system size.
Let the number of $2 \in B_1$ appearing in 
$b_1 \otimes \cdots \otimes b_L$ be $N$.
We assume $L \ge 2N$. The other case can be reduced to 
this case by the exchange $1 \leftrightarrow 2$ as
explained in $\S$3.3 of \cite{KTT}. 
Let ${\mathcal P}$ be the set of such states.
Then the time evolution 
$T_l: {\mathcal P} \rightarrow {\mathcal P}$
is defined by
\begin{equation}\label{eq:tl}
u_l \otimes p \simeq p^\ast \otimes v_l,\quad 
v_l \otimes p \simeq T_l(p)\otimes v_l.
\end{equation}
In the first relation, one applies the combinatorial $R$ 
$L$ times to carry $u_l$ through $p \in {\mathcal P}$ 
to the right. 
It determines $v_l \in B_l$ and 
$p^\ast \in {\mathcal P}$ uniquely. 
($p^\ast$ will play no role in the sequel.)
Then the second relation using the so obtained $v_l$ 
specifies $T_l(p)$, where 
the appearance of the same $v_l$ in the right hand side is a
non-trivial claim (\cite{KTT} (2.10)).
$v_l$ is dependent on $p$ as opposed to $u_l$.

The combinatorial $R$ is the identity map on $B_1 \otimes B_1$, therefore
$T_1$ is just the cyclic shift 
$T_1(b_1 \otimes \cdots \otimes b_L) = 
b_L \otimes b_1 \otimes \cdots \otimes b_{L-1}$.
The commutativity $T_lT_k = T_kT_l$ holds for any $k, l$
(\cite{KTT} Theorem 2.2).
\begin{example}\label{ex:t23}
We depict the relation $R(u \otimes b) = b' \otimes u'$ 
by a vertex diagram:
\begin{equation*}
\begin{picture}(90,40)(-20,-9)
\put(0,10){\line(1,0){20}}\put(-6,8){$u$}\put(22,8){$u'$}
\put(10,0){\line(0,1){20}}\put(8,25){$b$}\put(8,-9){$b'$}
\end{picture}
\end{equation*}
Then for $p=1121221 \in B_1^{\otimes 7}$ 
(symbol $\otimes$ omitted), 
one has $v_2=(1,1) = 12 \in B_2$ from (\ref{eq:tl}).
The time evolution 
$T_2(1121221)$ is obtained by the composition of the 
vertex diagrams

$$
\begin{picture}(160,35)(0,-18)

\multiput(0,0)(25,0){7}{
\put(0,0){\line(1,0){12}}
\put(6,-5){\line(0,1){10}}}

\put(3,9){1}
\put(28,9){1}
\put(53,9){2}\put(78,9){1}\put(103,9){2}\put(128,9){2}\put(153,9){1}

\put(-11,-3){12}\put(14,-3){11}
\put(39,-3){11}\put(64,-3){12}
\put(89,-3){11}\put(114,-3){12}
\put(139,-3){22}\put(164,-3){12}

\put(3,-14){2}
\put(28,-14){1}
\put(53,-14){1}\put(78,-14){2}\put(103,-14){1}
\put(128,-14){1}\put(153,-14){2}

\end{picture}
$$
leading to $T_2(1121221) = 2112112$.
\end{example}

\section{Combinatorial Bethe ansatz data}\label{sec:cb}

Let ${\mathcal P}_+$ be the subset of 
${\mathcal P}$ consisting of the $sl_2$-highest states.
Namely,
${\mathcal P}_+ = \{ p \in {\mathcal P} \mid {\tilde e}_1p = 0 \}$,
where ${\tilde e}_1$ is a Kashiwara operator \cite{Ka}.
In practice, $p=b_1 \otimes \cdots \otimes b_L$ is highest 
if and only if 
$b_1 \otimes \cdots \otimes b_k$ has nonnegative weights 
for any $1 \le k \le L$.

According to the combinatorial Bethe ansatz at $q=1$ \cite{KR},
there is one to one correspondence between 
${\mathcal P}_+$  and the set of the objects called 
the {\em rigged configurations}.
We let $\phi(p_+) = (\mu, J)$ denote 
the Kerov-Kirillov-Reshetikhin (KKR) bijection. 
Here $p_+$ is highest state 
and the right hand side is a rigged configuration.
The data $\mu$, called a configurations, is just 
a Young diagram for a partition of $N$, 
for which $N\le L/2$ is assumed.

In what follows, we let 
$\I = \{ i_1 < i_2 < \cdots < i_g \}$ be the set of 
distinct lengths of rows of $\mu$, and the multiplicity of 
the length $i (\in \I)$ rows by $m_i (\ge 1)$.
In terms of the non-increasing array of parts we have 
\begin{equation}\label{eq:mu}
\mu = (i_g^{m_{i_g}}\ldots i_1^{m_{i_1}}).
\end{equation}
Obviously, a configuration $\mu$ is equivalent with the data 
$\I$ and $m_{i_1}, \ldots, m_{i_g}$.

To each row of $\mu$, there is assigned an nonnegative 
integer called a rigging, which is represented 
by the symbol $J$ in $(\mu, J)$. See Example \ref{ex:45}.
Let $J_{i,1}, \ldots, J_{i,m_i}$  be the 
ones assigned to the $m_i$ rows of length $i$ in $\mu$
from the bottom to the top.
The data $(\mu,J)$ with 
$J =\{ (J_{i,\alpha}) \mid i \in \I, 1 \le \alpha \le m_i\}$ 
will be called a rigged configuration if 
$0 \le J_{i,1} \le \cdots \le J_{i, m_i} \le p_i$ holds for each
$i \in \I$. 
Here $p_i$ is defined by 
\begin{equation}\label{eq:vacancy}
p_i = L - 2\sum_{j \in \I}\min(i,j)m_j,
\end{equation}
which is nonnegative if $\mu$ is a configuration. 

We introduce the index set 
\begin{equation}
\tI= \{(i,\alpha) \mid i \in \I, 1 \le \alpha \le m_i \},
\quad \gamma = \vert \tI \vert = m_{i_1} + \cdots + m_{i_g}
\end{equation} 
labeling the rows of $\mu$.
The following vectors and matrix 
whose components are labeled with $\tI$ will be used:
\begin{align}
{\tilde {\bf h}}_j &= (\min(i,j))_{(i,\alpha) \in \tI} \in \Z^\gamma
\quad (j \in \I),\label{eq:htilde}\\
{\tilde {\bf J}} &= (J_{i,\alpha}+\alpha-1)_{(i,\alpha) \in \tI} \in \Z^\gamma,
\label{eq:jtilde}\\ 
A &= (A_{i\alpha, j\beta})_{(i,\alpha), (j,\beta) \in \tI}\;,\;\;
A_{i\alpha, j\beta} = \delta_{i,j}\delta_{\alpha, \beta}
(p_i+m_i) + 2\min(i,j) - \delta_{i,j}.\label{eq:A}
\end{align}
We also introduce the vectors and matrices
whose indices range over 
the previously introduced set $\I = \{i_1,\ldots, i_g\}$ as
\begin{align}
{\bf h}_j &= (\min(i,j))_{i\in \I} \in \Z^g,\label{eq:h}\\
{\bf p} &= (p_i)_{i \in \I} 
= L{\bf h}_1 - 2\sum_{j \in \I}m_j{\bf h}_j \in \Z^g,\label{eq:p}\\
{\bf J} & = (J_{i,1}+\cdots + J_{i,m_i})_{i \in \I} \in \Z^g,\label{eq:J}\\
F &=(F_{i,j})_{i,j \in \I},\;\;
F_{i,j} = \sum_{\beta=1}^{m_j}A_{i\alpha, j \beta} 
= \delta_{ij}p_i + 2\min(i,j)m_j, \label{eq:f}\\
M &= {\rm diag}(m_i)_{i \in \I},\label{eq:m}
\end{align}
The matrices $A, F$ were introduced 
in the study of $q=0$ Bethe equation \cite{KN}.
In particular, $A$ is the coefficient matrix of the 
linearized Bethe equation (string center equation),
which is known to be positive definite.
The result of the sum in (\ref{eq:f}) is independent of $\alpha$.

\section{Action-angle variable and inverse scattering transform}
\label{sec:ist}

Any state $p \in {\mathcal P}$ can be made highest 
by applying the cyclic shift appropriately.
Namely, one can express $p$ as $p = T_1^d(p_+)$ 
with some integer $d$ and highest state 
$p_+ \in {\mathcal P}_+$.
Although the choice of $d$ and $p_+$ is not unique,  
the configuration of $\phi(p_+)$ is unique.
The set of states is decomposed 
according to the configuration of $p_+$ as
${\mathcal P} = \sqcup_{\mu}{\mathcal P}(\mu)$. 
We call $\mu$ the action variable.
It is a conserved quantity (\cite{KTT}  Lemma C.3, Corollary 3.5), 
therefore each subset ${\mathcal P}(\mu)$ is invariant 
under any time evolution $T_l$.
The action variable $\mu$ is equivalent with the data 
$\I = \{i_1,\ldots, i_g\}$ and $\{m_i \mid i \in \I\}$ via (\ref{eq:mu}).
In the periodic box-ball system, 
$\I$ is the list of amplitudes of solitons and 
$m_i$ is the number of solitons with amplitude $i$.

Given an action variable $\mu$, 
the set of angle variable is given by 
\begin{equation}\label{eq:jv}
{\mathcal J}(\mu) = ({\mathcal I}_{m_{i_1}}\times 
\cdots \times {\mathcal I}_{m_{i_g}})/\Gamma,\quad 
\Gamma = A\Z^\gamma,
\end{equation}
where 
${\mathcal I}_n = (\Z^n-\Delta_n)/{\frak S}_n$
is the $n$-dimensional lattice without the ``diagonal points"
$\Delta_n = \{(z_1,\ldots, z_n) \in \Z^n
\mid z_\alpha = z_\beta \hbox{ for some }
1 \le \alpha \neq \beta \le  n \}$ divided by the 
action of the symmetric group ${\frak S}_n$.
Plainly, it is the set of array of $n$ distinct integers 
whose ordering does not matter.
In the sequel, elements in 
${\mathcal I}_{m_{i_1}}\times 
\cdots \times {\mathcal I}_{m_{i_g}}$ and its image in 
${\mathcal J}(\mu)$ will be denoted by the same symbol.
The set
${\mathcal J}(\mu)$ is finite 
and its cardinality is given by \cite{KN,KTT}
\begin{equation}\label{eq:card}
\vert {\mathcal J}(\mu)\vert = 
(\det F)\prod_{i \in \I} \frac{1}{m_i}
\binom{p_i + m_i - 1}{m_i - 1} = 
\frac{L}{p_{i_g}}
\prod_{i \in \I}\binom{p_i+m_i-1}{m_i}.
\end{equation}
This is a positive integer if 
$\mu$ is a configuration.
(When $\vert \mu\vert = L/2$, we have $p_{i_g} = 0$.
In this case, the factor 
$\frac{L}{p_{i_g}}\binom{p_{i_g}+m_{i_g}-1}{m_{i_g}}$ is to be 
understood as $\frac{L}{m_{i_g}}$.)
The formula (\ref{eq:card}) tells the number of states containing 
$m_i$ solitons with amplitude $i$.
The completeness of soliton states is known 
in the sense that $\sum_\mu \vert {\mathcal J}(\mu)\vert
= \binom{L}{N}$ for $0 \le N \le L/2$, where the sum extends
over those $m_i$ such that $\sum_iim_i = N$
(\cite{KN}, Theorems 3.5 and 4.9).
The states in section \ref{sec:intro} have the length $L=45$
and $\mu$ depicted in Example \ref{ex:45}.
For this $\mu$ and $L$, the formula (\ref{eq:card}) tells
$\vert {\mathcal J}(\mu) \vert = 316350$.

Let us specify the angle variable for a given state 
$p \in {\mathcal P}(\mu)$.
Suppose $p$ is expressed as $p = T_1^d(p_+)$
in terms of a highest state $p_+$ and the cyclic shift $T_1$.
Let the image of the KKR bijection be 
$\phi(p_+) = (\mu, J)$.
{}From the configuration $\mu$ we know
$\I = \{i_1 < i_2 < \cdots < i_g\}$ and 
$m_{i_1}, \ldots, m_{i_g}$ by (\ref{eq:mu}).
Then the angle variable is found by applying 
the definitions (\ref{eq:htilde}), (\ref{eq:jtilde}), (\ref{eq:A}) 
to $J=(J_{i,\alpha})_{(i,\alpha) \in \tI}$ as
\begin{equation}\label{eq:Phi}
\begin{split}
\Phi: \quad & {\mathcal P}(\mu)\;
\longrightarrow \;\Z \times {\mathcal P}_+
\longrightarrow \;\;\;\;
{\mathcal J}(\mu)\\
&\;\;p \;\quad\,\longmapsto \;(d, p_+) \,\longmapsto\;
({\tilde{\bf J}} + d{\tilde {\bf h}}_1)/\Gamma.
\end{split}
\end{equation}
Due to the shift $+\alpha-1$ in (\ref{eq:jtilde}) 
and the rigging condition 
$0 \le J_{i,1} \le \cdots \le J_{i, m_i} \le p_i$,
it follows that the components of ${\tilde{\bf J}}$ hence
${\tilde{\bf J}} + d{\tilde {\bf h}}_1$ within 
each block labeled by $i \in \I$ are distinct.
In (\ref{eq:Phi}) we regard the ${\tilde{\bf J}} + d{\tilde {\bf h}}_1$
as a representative of an element in ${\mathcal J}(\mu)$.
The non-uniqueness of $d, p_+$ 
is cancelled by ${\rm mod }\, \Gamma$ making 
$\Phi$ well-defined (\cite{KTT} Proposition 3.7).

\begin{theorem}[\cite{KTT} Theorem 3.12]\label{th:mae}
$\Phi$ is a bijection and the following diagram is commutative.
\begin{equation*}
\begin{CD}
{\mathcal P}(\mu) @>{\Phi}>> {\mathcal J}(\mu) \\
@V{T_l}VV @VV{T_l}V\\
{\mathcal P}(\mu) @>{\Phi}>> {\mathcal J}(\mu) 
\end{CD}
\end{equation*}
Here the time evolution on ${\mathcal J}(\mu)$ is defined by 
$T_l({\tilde {\bf J}}) = {\tilde {\bf J}} + {\tilde {\bf h}}_l$.
\end{theorem}

\vspace{0.3cm}
The composition $\Phi^{-1}\circ T_l \circ \Phi$ 
achieves the solution of the initial value problem 
by the inverse scattering transform.
Namely, $\Phi^{\pm 1}$ is the direct/inverse scattering map which 
transforms the dynamics $T_l$ of the periodic box-ball system into 
the straight motion on ${\mathcal J}(\mu)$ with the velocity
${\tilde {\bf h}}_l$.
In this sense, ${\mathcal J}(\mu)$ serves as an ultradiscrete analogue 
of the Jacobi variety in quasi-periodic solutions to soliton equations
\cite{DT,DMN,KM}.

By using Theorem \ref{th:mae}, a closed formula for 
the fundamental period, i.e., the
smallest positive integer such that $T^{\mathcal N}_l(p) = p$ 
for any $l$ and $p$ 
has been obtained in Theorem 4.9 of \cite{KTT}, which 
supplements a recursive description 
of the $l=\infty$ case \cite{YYT}.
The states in section \ref{sec:intro} have the fundamental period $3515$. 
See also \cite{KT} for a generalization to $A^{(1)}_n$ based on the 
Bethe eigenvalue at $q=0$. 
 
We note that the time evolution of the angle variable
${\tilde {\bf J}}$ (\ref{eq:jtilde}) leads to the rule 
\begin{equation}\label{eq:tj}
T_l({\bf J}) = {\bf J} + M{\bf h}_l
\end{equation}
for the ``bundled" angle variable ${\bf J}$ (\ref{eq:J}),
where $M$ is defined by (\ref{eq:m}).

\section{Explicit formula for the solution of the initial value problem}
\label{sec:ivp}

Fix the action variable $\mu$ (\ref{eq:mu}) and set 
\begin{equation}\label{eq:omega}
\Omega = MF = (\Omega_{i,j})_{i,j \in \I},
\;\;
\Omega_{i,j} = \delta_{ij}p_im_i+2\min(i,j)m_im_j,
\end{equation}
where $F$ is defined in (\ref{eq:m}).
$\Omega$ is a positive definite symmetric integer matrix
which will play the role of the period matrix.
Note that 
\begin{equation}\label{eq:ao}
\Omega {\bf h}_1 = LM{\bf h}_1.
\end{equation}

In addition to the ultradiscrete Riemann theta function 
$\Theta_{\bf a}({\bf z})$ in (\ref{eq:urt}),
we need $\chi({\bf s};{\tilde{\bf I}})$.
It is a function of ${\bf s} = (s_i)_{i \in \I} \in \Z^g$ as well as
${\tilde{\bf I}} = 
(I_{i,\alpha}+\alpha-1)_{(i,\alpha) \in \tI} \in 
{\mathcal I}_{m_{i_1}}\times 
\cdots \times {\mathcal I}_{m_{i_g}}$ that takes rational values:
\begin{align}
\chi({\bf s};{\tilde{\bf I}}) &= 
\chi({\bf s}';{\tilde{\bf I}})\;\; \hbox{if } {\bf s} \equiv {\bf s}' 
\mod M\Z^g,\\
\chi({\bf s}; {\tilde{\bf I}}) &= \sum_{i \in \I}\frac{1}{m_i}
\sum_{1 \le \alpha \le s_i < \beta \le m_i}
(I_{i,\beta}-I_{i,\alpha}-\frac{p_i}{2})\label{eq:chi}\\
&\hbox{if } \, 0 \le s_i < m_i, \quad
I_{i,1}\le \cdots \le I_{i,m_i}\, \hbox{ for all } \, i \in \I.\nonumber
\end{align}
Here the second sum in (\ref{eq:chi}) 
extends over the pairs $(\alpha, \beta)$.
In particular, $\chi({\bf 0};{\tilde{\bf I}}) = 0$.
The condition $I_{i,1}\le \cdots \le I_{i,m_i}$ is needed 
to cope with the ${\mathfrak S}_{m_i}$ symmetry 
in ${\mathcal I}_{m_i}$.

Given 
${\bf I} \in \Z^g$ and 
${\tilde{\bf I}} = 
(I_{i,\alpha}+\alpha-1)_{(i,\alpha) \in \tI} \in 
{\mathcal I}_{m_{i_1}}\times 
\cdots \times {\mathcal I}_{m_{i_g}}$, 
we define the ultradiscrete tau function by
\begin{equation}\label{eq:tau}
\tau_r(k) = \max_{{\bf s} \in \Z^g/M\Z^g}\bigl\{
\Theta_{M^{-1}{\bf s}}
\Bigl({\bf I} + M\bigl(r{\bf h}_\infty-k{\bf h}_1
- \frac{\bf p}{2}\bigr)\Bigr) + \chi({\bf s}; {\tilde{\bf I}}) \bigr\}
\end{equation}
for $r\in \{0,1\},\; k \in \Z$.
Here ${\bf h}_j$, ${\bf p}$ and $M$ are those in 
(\ref{eq:h}), (\ref{eq:p}) and (\ref{eq:m}).

Let us specify ${\tilde{\bf I}}$ and ${\bf I}$ in (\ref{eq:tau})
from a given state $p \in {\mathcal P}(\mu)$.
As for ${\tilde{\bf I}}$, it is taken to be the 
angle variable of $p \in {\mathcal P}(\mu)$ 
according to (\ref{eq:Phi}),
and ${\bf I}$ is determined from it as follows:
\begin{align}
{\tilde{\bf I}} &= \Phi(p) =
(I_{i,\alpha}+\alpha-1)_{(i,\alpha) \in \tI} \in {\mathcal J}(\mu),
\label{eq:Idef}\\
{\bf I} &= (I_{i,1}+\cdots + I_{i,m_i})_{i \in \I} \in \Z^g/F\Z^g.
\label{eq:Idef2}
\end{align}
This is the same relation as 
(\ref{eq:jtilde}) and (\ref{eq:J}).
Note that the elements ${\tilde{\bf I}}$ and ${\bf I}$
are defined mod $\Gamma=A\Z^\gamma$ and mod $F\Z^g$,
respectively, which is consistent with the relation (\ref{eq:f}) between 
$A$ and $F$. Now we state the main result of this paper.

\begin{theorem}\label{th:ima}
The state 
$p \in {\mathcal P}(\mu)$ corresponding to the action variable
$\mu$ and the angle variable ${\tilde{\bf I}}$ is expressed as 
$p = (1-x_1,x_1)\otimes \cdots \otimes (1-x_L, x_L)$, where 
$x_k=0,1$ is given by
\begin{equation}\label{eq:main}
x_k  = \tau_0(k)-\tau_0(k-1)-\tau_1(k)+\tau_1(k-1).
\end{equation}
\end{theorem}

The proof is similar to section 3 in \cite{KS}.
Combined with Theorem \ref{th:mae}, 
Theorem \ref{th:ima} leads to an explicit formula for 
the solution of the initial value problem.
The state $(\prod_lT^{c_l}_l)(p)$ is obtained just by 
replacing the angle variable as
\begin{equation}\label{eq:clt}
{\tilde{\bf I}} \mapsto {\tilde{\bf I}}  + \sum_l c_l {\tilde{\bf h}}_l
\end{equation}
in (\ref{eq:tau}).
Under the change (\ref{eq:clt}), the function
$\chi({\bf s}; {\tilde{\bf I}})$ (\ref{eq:chi}) is invariant
since it only depends on the difference $I_{i,\beta}-I_{i,\alpha}$.
On the other hand, the variable ${\bf I}$ entering (\ref{eq:tau}) 
changes as 
\begin{equation}\label{eq:evo}
{\bf I} \mapsto {\bf I} + M(\sum_lc_l{\bf h}_l)
\end{equation}
according to the rule  (\ref{eq:tj}).
Especially, $T_1^L({\bf I}) = {\bf I} + LM{\bf h}_1 
= {\bf I} + \Omega{\bf h}_1$ by 
(\ref{eq:ao}).
Thus the tau function only exhibits the change 
due to the quasi-periodicity (\ref{eq:qp}) leaving 
(\ref{eq:main}) invariant.
This is consistent with the fact that the cyclic shift 
$T_1^L$ is trivial and 
(\ref{eq:main}) is a period $L$ function of $k$.
We remark that our tau function 
(\ref{eq:tau}) is always integer-valued.

{}From (\ref{eq:h}) we see 
${\bf h}_l={\bf h}_\infty$ if $l \ge \max \mathbb{I} \,(=i_g)$.
Therefore the effect (\ref{eq:evo}) of the time evolution $T_l$ 
is ${\bf I} \mapsto {\bf I}+M{\bf h}_\infty$, which is 
equivalent to setting $r \mapsto r+1$ in (\ref{eq:tau}).
In other words, $r$ serves as the time variable for such $T_l$, i.e.,
$T_l^t(p)=(1-x_{1,t},x_{1,t})\otimes \cdots \otimes 
(1-x_{L,t}, x_{L,t})$ 
with 
\begin{equation}\label{eq:add}
x_{k,t}= \tau_t(k)-\tau_t(k-1)-\tau_{t+1}(k)+\tau_{t+1}(k-1)
\end{equation}
holds for $l \ge \max \mathbb{I}$.

\begin{example}\label{ex:45}
Let us consider the $t=0,1$ states in section \ref{sec:intro}:
\begin{equation}\label{eq:jotai}
\begin{split}
p &= 111222221111222222221111111111111112222111211\\
T_9(p) &= 111111112222111111112222222221111111111222122,
\end{split}
\end{equation}
which are elements in $B_1^{\otimes 45}$.
We have $p=T_1^{29}(p_+)$
with a highest state 
$$
p_+ = 111111222211121111122222111122222222111111111
$$
which corresponds to the rigged configuration:
$$
\begin{picture}(100,56)(0,-10)
\put(0,40){\line(1,0){90}} \put(95,31){0}
\put(0,30){\line(1,0){90}} \put(43,21){6}
\put(0,20){\line(1,0){40}} \put(43,11){2}
\put(0,10){\line(1,0){40}} \put(13,1){10}
\put(0,0){\line(1,0){10}}

\put(0,0){\line(0,1){40}}
\put(10,0){\line(0,1){40}}
\put(20,10){\line(0,1){30}}
\put(30,10){\line(0,1){30}}
\put(40,10){\line(0,1){30}}
\put(50,30){\line(0,1){10}}
\put(60,30){\line(0,1){10}}
\put(70,30){\line(0,1){10}}
\put(80,30){\line(0,1){10}}
\put(90,30){\line(0,1){10}}

\end{picture}
$$
Thus $\I=\{1,4,9\}, \;M={\rm diag}(m_1,m_4,m_9)={\rm diag}(1,2,1)$.
$I_{i,\alpha}$ in (\ref{eq:Idef}) reads
$(I_{1,1},I_{4,1},I_{4,2},I_{9,1}) = (39,31,35,29)$.
The data ${\bf p}$ (\ref{eq:p}), 
${\bf I}$  (\ref{eq:Idef2}) and $\Omega$ (\ref{eq:omega}) are given by
\begin{align*}
{\bf p} = \begin{pmatrix}p_1 \\ p_4 \\ p_9\end{pmatrix}
=\begin{pmatrix}37 \\ 19 \\ 9\end{pmatrix},\;\;
{\bf I} = \begin{pmatrix}39 \\ 66 \\ 29\end{pmatrix},\;\;
\Omega = \begin{pmatrix}
39 & 4 & 2\\ 4 & 70 & 16 \\ 2 & 16 & 27\end{pmatrix}.
\end{align*}
The argument of the tau function (\ref{eq:tau}) reads
\begin{equation}\label{eq:zrk}
{\bf I} + M\bigl(r{\bf h}_\infty-k{\bf h}_1
- \frac{\bf p}{2}) = 
\begin{pmatrix}r-k+41/2 \\ 8r-2k+47 \\ 9r-k+49/2\end{pmatrix}
=:{\bf z}_{r,k}\;.
\end{equation}

The function $\chi({\bf s};\tI)$ in (\ref{eq:chi}) is 
non-vanishing at ${\bf s}=(0,1,0)$ taking the value
$\frac{1}{m_4}(I_{4,2}-I_{4,1}-\frac{p_4}{2}) = -\frac{11}{4}$.
Thus the tau function is given by
$$
\tau_r(k) = \max\{\Theta_{0,0,0}({\bf z}_{r,k}), \;
\Theta_{0,\frac{1}{2},0}({\bf z}_{r,k})-\frac{11}{4}\}.
$$
The time evolution pattern in section \ref{sec:intro}
is reproduced according to (\ref{eq:add}). 
Here is the plot of $\tau_r(k)$ for 
$0 \le r \le 2$ and $0 \le k \le 45$.

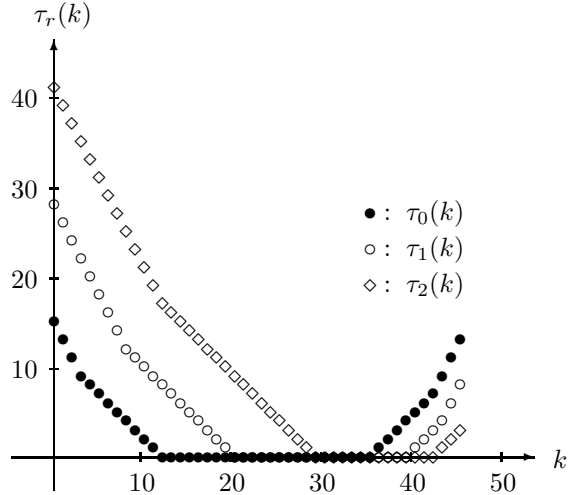
\begin{equation*}
\setlength{\unitlength}{1.2mm}
\begin{picture}(50,58)(0,-5)

\put(-3,0.7){\vector(1,0){58}} \put(57,-0.5){$k$}
\put(1.7,-3){\vector(0,1){50}} \put(-0.5,49){$\tau_r(k)$}

\put(1.2,10.5){\line(1,0){1}}\put(-3,9.5){10}
\put(1.2,20.5){\line(1,0){1}}\put(-3,19.5){20}
\put(1.2,30.5){\line(1,0){1}}\put(-3,29.5){30}
\put(1.2,40.5){\line(1,0){1}}\put(-3,39.5){40}

\put(11.3,0.2){\line(0,1){1}}\put(10,-3){10}
\put(21.3,0.2){\line(0,1){1}}\put(20,-3){20}
\put(31.3,0.2){\line(0,1){1}}\put(30,-3){30}
\put(41.3,0.2){\line(0,1){1}}\put(40,-3){40}
\put(51.3,0.2){\line(0,1){1}}\put(50,-3){50}

\put(36,27){$\bullet: \;\tau_0(k)$}
\put(36,23){$\circ: \; \tau_1(k)$}
\put(36,19){$\diamond: \; \tau_2(k)$}

\put(1, 15){$\bullet$} \put(2, 13){$\bullet$} \put(3, 11){$\bullet$} \put(4, 9){$\bullet$} 
\put(5, 8){$\bullet$} 
\put(6, 7){$\bullet$} \put(7, 6){$\bullet$} \put(8, 5){$\bullet$} \put(9, 4){$\bullet$} 
\put(10, 3){$\bullet$} \put(11, 2){$\bullet$} \put(12, 1){$\bullet$} \put(13, 0){$\bullet$} 
\put(14, 0){$\bullet$} 
\put(15, 0){$\bullet$} \put(16, 0){$\bullet$} \put(17, 0){$\bullet$} \put(18, 0){$\bullet$} 
\put(19, 0){$\bullet$} \put(20, 0){$\bullet$} \put(21, 0){$\bullet$} \put(22, 0){$\bullet$} 
\put(23, 0){$\bullet$} 
\put(24, 0){$\bullet$} \put(25, 0){$\bullet$} \put(26, 0){$\bullet$} \put(27, 0){$\bullet$} 
\put(28, 0){$\bullet$} \put(29, 0){$\bullet$} \put(30, 0){$\bullet$} \put(31, 0){$\bullet$} 
\put(32, 0){$\bullet$} \put(33, 0){$\bullet$} \put(34, 0){$\bullet$} \put(35, 0){$\bullet$} 
\put(36, 0){$\bullet$} \put(37, 1){$\bullet$} \put(38, 2){$\bullet$} \put(39, 3){$\bullet$} 
\put(40, 4){$\bullet$} \put(41, 5){$\bullet$} 
\put(42, 6){$\bullet$} \put(43, 7){$\bullet$} \put(44, 9){$\bullet$} \put(45, 11){$\bullet$} 
\put(46, 13){$\bullet$} 

\put(1, 28){$\circ$} \put(2, 26){$\circ$} \put(3, 24){$\circ$} \put(4, 22){$\circ$} 
\put(5, 20){$\circ$} \put(6, 18){$\circ$} \put(7, 16){$\circ$} \put(8, 14){$\circ$} 
\put(9, 12){$\circ$} 
\put(10, 11){$\circ$} \put(11, 10){$\circ$} \put(12, 9){$\circ$} \put(13, 8){$\circ$} 
\put(14, 7){$\circ$} \put(15, 6){$\circ$} \put(16, 5){$\circ$} \put(17, 4){$\circ$} 
\put(18, 3){$\circ$} \put(19, 2){$\circ$} \put(20, 1){$\circ$} \put(21, 0){$\circ$} 
\put(22, 0){$\circ$} 
\put(23, 0){$\circ$} \put(24, 0){$\circ$} \put(25, 0){$\circ$} \put(26, 0){$\circ$} 
\put(27, 0){$\circ$} \put(28, 0){$\circ$} \put(29, 0){$\circ$} \put(30, 0){$\circ$} 
\put(31, 0){$\circ$} 
\put(32, 0){$\circ$} \put(33, 0){$\circ$} \put(34, 0){$\circ$} \put(35, 0){$\circ$} 
\put(36, 0){$\circ$} \put(37, 0){$\circ$} \put(38, 0){$\circ$} \put(39, 0){$\circ$} 
\put(40, 0){$\circ$} 
\put(41, 1){$\circ$} \put(42, 2){$\circ$} \put(43, 3){$\circ$} \put(44, 4){$\circ$} 
\put(45, 6){$\circ$} \put(46, 8){$\circ$}

\put(1, 41){$\diamond$} \put(2, 39){$\diamond$} \put(3, 37){$\diamond$} \put(4, 35){$\diamond$} 
\put(5, 33){$\diamond$} \put(6, 31){$\diamond$} \put(7, 29){$\diamond$} \put(8, 27){$\diamond$} 
\put(9, 25){$\diamond$} 
\put(10, 23){$\diamond$} \put(11, 21){$\diamond$} \put(12, 19){$\diamond$} \put(13, 17){$\diamond$} 
\put(14, 16){$\diamond$} \put(15, 15){$\diamond$} \put(16, 14){$\diamond$} \put(17, 13){$\diamond$} 
\put(18, 12){$\diamond$} \put(19, 11){$\diamond$} \put(20, 10){$\diamond$} \put(21, 9){$\diamond$} 
\put(22, 8){$\diamond$} 
\put(23, 7){$\diamond$} \put(24, 6){$\diamond$} \put(25, 5){$\diamond$} \put(26, 4){$\diamond$} 
\put(27, 3){$\diamond$} \put(28, 2){$\diamond$} \put(29, 1){$\diamond$} \put(30, 0){$\diamond$} 
\put(31, 0){$\diamond$} 
\put(32, 0){$\diamond$} \put(33, 0){$\diamond$} \put(34, 0){$\diamond$} \put(35, 0){$\diamond$} 
\put(36, 0){$\diamond$} \put(37, 0){$\diamond$} \put(38, 0){$\diamond$} \put(39, 0){$\diamond$} 
\put(40, 0){$\diamond$} 
\put(41, 0){$\diamond$} \put(42, 0){$\diamond$} \put(43, 0){$\diamond$} \put(44, 1){$\diamond$} 
\put(45, 2){$\diamond$} \put(46, 3){$\diamond$}

\end{picture}
\end{equation*}
The boundary values are $(\tau_0(0),\tau_1(0),\tau_2(0))=(15,28,41)$
and $(\tau_0(45),\tau_1(45),\tau_2(45))=(13,8,3)$.
In the plot of $\tau_t(k)$ or $\tau_{t+1}(k)$,
one observes a change of the gradient 
at those $k$ corresponding to the front or tail 
of solitons in $T^t_9(p)$ (\ref{eq:jotai}).
\end{example}

\vspace{0.2cm}
The tau function (\ref{eq:tau}) simplifies considerably 
if there are no solitons with the same amplitude, namely,
$\forall m_i=1$.
We have $\chi({\bf s};{\tilde{\bf I}}) = 0$, $\tI=\I$ and  
$\Omega = F = (\delta_{ij}p_i+2\min(i,j))_{i,j \in \I}$.
Let us write the ultradiscrete Riemann theta function with 
characteristics ${\bf 0}$ as
$\Theta({\bf z}) = \Theta_{\bf 0}({\bf z})$.
Then $\tau_r(k) = 
\Theta\bigl({\bf I}-\frac{\bf p}{2}-k{\bf h}_1+r{\bf h}_\infty\bigr)$.
Consequently (\ref{eq:main}) reduces to 
\begin{equation*}
\begin{split}
x_k &= \Theta\bigl(
{\bf I}-\frac{\bf p}{2}-k{\bf h}_1\bigr) - 
\Theta\bigl(
{\bf I}-\frac{\bf p}{2}-(k\!-\!1){\bf h}_1\bigr)\\
&-\Theta\bigl(
{\bf I}-\frac{\bf p}{2}-k{\bf h}_1+{\bf h}_\infty\bigr)+
\Theta\bigl(
{\bf I}-\frac{\bf p}{2}-(k\!-\!1){\bf h}_1+{\bf h}_\infty\bigr),
\end{split}
\end{equation*}
which was obtained in \cite{KS}.
This has the same structure as the ultradiscretization of the
solution of the periodic Toda lattice 
by Date and Tanaka \cite{DT}.

\vspace{0.3cm}\noindent
{\bf Acknowledgments} \hspace{0.1cm}
The authors thank Rei Inoue for a discussion.
R.S. is a research fellow of the Japan Society 
for the Promotion of Science.
He thanks Miki Wadati for continuous encouragements.

\vspace{5mm}
\begin{flushleft}
Atsuo Kuniba:\\
Institute of Physics, Graduate School of Arts and Sciences,
University of Tokyo,
Komaba, Tokyo 153-8902, Japan\\
\texttt{atsuo@gokutan.c.u-tokyo.ac.jp}\vspace{3mm}\\
Reiho Sakamoto:\\
Department of Physics, Graduate School of Science, 
University of Tokyo, Hongo, Tokyo 113-0033, Japan\\
\texttt{reiho@monet.phys.s.u-tokyo.ac.jp}
\end{flushleft}

\begin{thebibliography}{A}


\bibitem{AS}
Ablowitz, M. J. and Segur, H.: 
Solitons and the inverse scattering transform,
SIAM Studies in Appl. Math. 4. Philadelphia Pa. (1981)

\bibitem{Ba}
Baxter, R. J.:
Exactly solved models in statistical mechanics, 
Academic Press, London (1982)

\bibitem{Be}
Bethe, H. A.:
Zur Theorie der Metalle, I. Eigenwerte und
Eigenfunktionen der linearen Atomkette,
Z.\ Physik {\bf 71}  205--231 (1931)

\bibitem{DT}
Date, E. and Tanaka, S.: 
Periodic multi-soliton solutions of Korteweg-de Vries equation 
and Toda lattice,
Prog. Theoret. Phys. Suppl.  {\bf 59}  107--125 (1976)

\bibitem{DMN}
Dubrovin, B. A.,  Matveev, V. B. and Novikov, S. P.:
Nonlinear equations of Korteweg-de Vries type, 
finite-band linear operators and Abelian varieties
Russian Math. Surveys {\bf 31} 59--146 (1976) 

\bibitem{FOY}
Fukuda, K., Okado, M. and Yamada, Y.: 
{Energy functions in box ball systems},
Int.\ J.\ Mod.\ Phys.\ A {\bf 15} 1379--1392 (2000) 

\bibitem{GGKM}
Gardner, C. S., Greene, J. M., Kruskal, M. D. and Miura, R. M.:
Method for solving the Korteweg-de Vries equation,
Phys. Rev. Lett. {\bf 19} 1095--1097 (1967) 


\bibitem{HHIKTT}
Hatayama, G., Hikami, K., Inoue, R., Kuniba, A., Takagi T. and Tokihiro, T.:
The $A^{(1)}_M$ automata related to crystals of symmetric tensors,
{\it J.\ Math.\ Phys}. \ {\bf 42} 274--308 (2001) 

\bibitem{HKT} 
Hatayama, G., Kuniba, A. and Takagi, T.:
Soliton cellular automata associated with crystal bases,
Nucl. Phys. B{\bf 577}[PM] 619--645 (2000) 


\bibitem{JM}
Jimbo, M. and Miwa, T.:
Solitons and infinite dimensional Lie algebras,
Publ. RIMS. Kyoto Univ. {\bf 19} 943--1001 (1983)


\bibitem{KM}
Kac, M. and Moerbeke, P.:
A complete solution of the periodic Toda problem,
Proc. Nat. Acad. Sci. USA. {\bf 72} 2879--2880 (1975)


\bibitem{Ka}
Kashiwara, M.:
On crystal bases of the $q$-analogue of universal enveloping algebras.
Duke Math. J. {\bf 63} 465--516 (1991)


\bibitem{KR}
Kirillov, A. N. and Reshetikhin, N. Yu.:
The Bethe ansatz and the combinatorics of Young tableaux,
J. Soviet Math. {\bf 41}  925--955 (1988)

\bibitem{KN}
Kuniba, A. and Nakanishi, T.: 
The Bethe equation at $q=0$, the M\"obius inversion formula, 
and weight multiplicities: I. 
The $sl(2)$ case, Prog. in Math. {\bf 191} 185--216 (2000) 

\bibitem{KS}
Kuniba, A.  and Sakamoto, R.:
The Bethe ansatz in a periodic box-ball system
and the ultradiscrete Riemann theta function,
J. Stat. Mech.  P09005 (2006)

\bibitem{KSY}
Kuniba, A. , Sakamoto, R. and Yamada, Y.:
Tau functions in combinatorial Bethe ansatz,
preprint math.QA/0610505 

\bibitem{KTT} 
Kuniba, A., Takagi, T. and Takenouchi, A.:
Bethe ansatz and inverse scattering transform 
in a periodic box-ball system, 
Nucl. Phys. B [PM] 354--397 (2006)

\bibitem{KT}
Kuniba, A. and Takenouchi, A.:
Bethe ansatz at $q=0$ and periodic box-ball systems,
J. Phys. A: Math. Gen.  {\bf 39}  2551--2562 (2006)

\bibitem{M} 
Mumford, D.:
Tata Lectures on Theta II, Birkh\"auser, Boston (1984)


\bibitem{TS} Takahashi, D. and Satsuma, J.:
A soliton cellular automaton,
J. Phys. Soc. Jpn. {\bf 59}  3514--3519 (1990)


\bibitem{TTMS} 
Tokihiro, T., Takahashi, D., Matsukidaira and J. Satsuma, J.:
From soliton equations to integrable cellular automata through
a limiting procedure,
Phys. Rev. Lett. {\bf 76}  3247--3250 (1996)


\bibitem{YYT}
Yoshihara, D., Yura, F.  and Tokihiro, T.:
Fundamental cycle of a periodic box-ball system,
J. Phys. A: Math. Gen. {\bf 36}  99--121 (2003)

\end{thebibliography}
\end{document}